\documentclass[10pt, conference]{IEEEtran}

\usepackage{flushend}
\usepackage{tabularx}
\usepackage{amssymb}
\usepackage{wrapfig}
\usepackage[bookmarks=false]{hyperref}
\usepackage{rotating}
\usepackage{xspace}
\usepackage{multirow}
\usepackage{array}
\usepackage{amssymb}
\usepackage{amsfonts}
\usepackage{graphicx}
\usepackage{epstopdf}
\usepackage{amsmath}
\usepackage{booktabs}
\usepackage{epigraph}
\usepackage{listings}
\usepackage{subcaption}
\usepackage[table]{xcolor}
\usepackage[utf8]{inputenc}
\usepackage[normalem]{ulem} 
\usepackage{cleveref}
\usepackage{diagbox}
\crefname{section}{§}{§§}
\Crefname{section}{§}{§§}
\usepackage{tikz}
\usepackage{verbatim}
\usepackage{float}

\usepackage[normalem]{ulem}
\usepackage{color, colortbl}
\definecolor{Gray}{gray}{0.9}
\definecolor{White}{rgb}{255,255,255}

\newcommand*\circled[1]{\tikz[baseline=(char.base)]{
            \node[shape=circle,draw,inner sep=2pt] (char) {#1};}}

\long\def\responses{72\xspace}

\newcommand{\MyPara}[1]{\vspace{.2em}\noindent\textit{\textbf{#1}}\hspace{.3em}}

\newcommand\blfootnote[1]{%
  \begingroup
  \renewcommand\thefootnote{}\footnote{#1}%
  \addtocounter{footnote}{-1}%
  \endgroup
}

\begin{document}
% Title Suggestion: "Bugs fixes at a scientific software company"
%Does Background Matter? Measuring the Quality of Scientific Software
\title{Characterizing the Roles of Contributors in Open-source Scientific Software Projects}

\author{\IEEEauthorblockN{Reed Milewicz}
\IEEEauthorblockA{Sandia National Laboratories\\
Email: rmilewi@sandia.gov}
\and
\IEEEauthorblockN{Gustavo Pinto}
\IEEEauthorblockA{Federal University of Par\'{a}\\
Email: gpinto@ufpa.br}
\and
\IEEEauthorblockN{Paige Rodeghero}
\IEEEauthorblockA{Clemson University\\
Email: prodegh@clemson.edu}
}

% make the title area
\maketitle
\begin{abstract}

The development of scientific software is, more than ever, critical to the practice of science, and this is accompanied by a trend towards more open and collaborative efforts. Unfortunately, there has been little investigation into who is driving the evolution of such scientific software or how the collaboration happens. In this paper, we address this problem. We present an extensive analysis of seven open-source scientific software projects in order to develop an
empirically-informed model of the development process. This analysis was complemented by a survey of \responses scientific software developers. In the majority of the projects, we found senior research staff (e.g. professors) to be responsible for half or more of commits (an average commit share of 72\%) and heavily involved in architectural concerns (seniors were more likely to interact with files related to the build system, project meta-data, and developer documentation). Juniors (e.g. graduate students) also contribute substantially --- in one studied project, juniors made almost 100\% of its commits. Still, graduate students had the longest contribution periods among juniors (with 1.72 years of commit activity compared to 0.98 years for postdocs and 4 months for undergraduates). Moreover, we also found that third-party contributors are scarce, contributing for just one day for the project. 
%While we observed that senior members are generally more active than juniors, juniors play a pivotal role in creating and nurturing new software features necessary for research. 
The results from this study aim to help scientists to better understand their own projects, communities, and the contributors' behavior, while paving the road for future software engineering research.

%We also observed that senior members are more active than juniors, when considering different metrics such as file ownership or tests creation. Nevertheless, we also found that juniors play a pivotal role on XXX. The results from the study can be useful for helping scientists in universities or labs to better understand the hidden challenges of developing scientific software.

\end{abstract}

\blfootnote{Sandia National Laboratories is a multimission laboratory managed and operated by National Technology \& Engineering Solutions of Sandia, LLC, a wholly owned subsidiary of Honeywell International Inc., for the U.S. Department of Energy's National Nuclear Security Administration under contract DE-NA0003525. SAND2018-9345C.}

% no keywords

%software repository mining, scientific software

\IEEEpeerreviewmaketitle

\section{Introduction}

Computing technologies have had a profound impact on the practice of science: simulation and data-intensive computation are now known as the third and fourth paradigms of science, on equal footing with experimentation and theory~\cite{hey2009fourth}. This shift has accelerated the growth of a diverse ecosystem of scientific software projects. The term ``scientific software'' is an umbrella that covers all aspects of the research pipeline, including codes for simulation and data analysis, dataset management, communication infrastructure, and underlying mathematical libraries~\cite{carver2016software}. It is software that exists ``to support the exploration of a scientific question''~\cite{mesh2013scientific}.

%A unifying definition for this broad class of software, as expressed in Mesh and Hawker, is a product that exists ``to support the exploration of a scientific question''~\cite{mesh2013scientific}.

What makes scientific software projects different from traditional software projects? Scientific software operates at the boundaries of human knowledge and tends to be in constant flux as new insights motivate unforeseen changes in requirements~\cite{letondal2003anticipating}. As noted by Segal~\cite{segal2008scientists}, this pressing need to produce or enable the production of knowledge lends itself to a mindset where ``software is valued only insofar as it progresses the science'', often in conflict with the need to have reliable, maintainable code. However, Turk and colleagues remarked that, in an era of increasing scale and complexity, ``the cyber-infrastructure necessary to address problems in computational science is no longer tractably solved by individuals working in isolation''~\cite{turk2013scaling}; broader, more open collaboration necessitates a shift in how the software is developed. From a software engineering research perspective, this motivates important questions about how the software evolves, who develops it, and how quality can emerge from this process.

%From a software engineering research perspective, this motivates important questions about how the software evolves, who develops it, and how quality can emerge from this process. Unfortunately, obtaining the data needed to answer these questions is not easy. Conti and Liu~\cite{conti2015bringing}, when studying the personnel composition at the MIT Department of Biology, argue that a lack of fine-grained data inhibits the study of research environments. This is also true of the software itself, in part because software provides a competitive advantage (incentivizing closed-source development). Perhaps just as importantly, unlike most modern, traditional open-source projects, common wisdom has it that an alarming number of scientific software projects eschew the use of version control systems; one domain scientist relayed to the authors frustrating experiences with projects relying on email chains and \texttt{tar.gz} attachments to manage their source code. This is confirmed by Nyugen-Hoan et al.'s survey~\cite{nguyen2010survey}, which found only half of respondents used a version control system in their software development. Fortunately, this trend is steadily changing and an increasing number of projects are adopting open practices (source, data, and access), providing us with more data and new opportunities for investigation.

 %whom Lee et al. refers to as the ``human infrastructure of cyber-infrastructure''~\cite{lee2006human}. 
We focus on the people meeting the demand for scientific software. Such scientific software developers represent a population so far not properly understood, since their characteristics, motivation, and needs to contribute to scientific software projects are intrinsically different than what drives traditional open source contributors. For instance, the actors that play the scientific developer role include students, postdocs, faculty, and staff. Their knowledge, skills, and goals can vary greatly, while also contributing to projects in different ways throughout their tenure. As a consequence, the plethora of existing studies on open source contributors might not help much, since they hardly take into account their roles or the complexity of the domains that scientific software is immersed in.

Much is still unknown about the state-of-the-practice of developing scientific software. For instance, who performs the majority of commit activities? Who fixes bugs? In order to better understand the relationship between these contributors and the software, %instead of asking scientific software developers their perceptions, 
we first leverage the availability and transparency of social coding websites to inspect data related to source code contributions and contributors. We selected a curated list of seven open-source scientific software projects by searching three different platforms: the Journal of Open Source Software, GitHub, and DOECODE, a platform for publicly funded DOE research codes. For each selected project, we identified the roles played by different contributors by analyzing each projects' documentation, websites, and other readily available sources. We then surveyed representative scientific software developers in order to cross-validate the findings found via the repositories' analysis.
%We use software repository mining techniques to model their commit histories, allowing us to qualitatively study the character of their contributions.

Using quantitative and qualitative data, our study produced a set of findings, some of which confirmed anecdotal accounts while others were unexpected. We discuss them in detail in Section~\ref{sec:analysis}. In the following, we highlight three of them.

%However, to our knowledge, no one has actually conducted a study on the work that different kinds of contributors do and the influence that they have. 

%In order to better understand the relationship between these contributors and the software, we present an analysis of multiple open-source scientific software projects. For each project, we identified the roles played by different contributors by analyzing documentation, project websites, and other readily available sources. Armed with this information, we used software repository mining techniques to model their commit histories, allowing us to qualitatively study the character of their contributions.

\begin{itemize}

\item \textbf{Senior researchers tend to be the most active and prolific contributors in terms of commits and file creation.} In four of the seven projects we studied, faculty and staff contributors were responsible for half or more of commits made to the project (with an average commit share of 72\%). In five projects, senior members were also responsible for the majority of files created and, by that measure, the resulting project structure. This influence over the overall direction of the software project was also evident in the fact that senior researchers were the most likely to have interacted with files related to the build system, project metadata, and developer documentation.

\item \textbf{Junior contributors, especially graduate students, are critical drivers of new features as well as supporting activities like test creation}. On average, junior contributors were responsible for 42\% of commits across all projects we studied; in one case, juniors were responsible for nearly 100\% of all commit activity. The majority of these commits came from graduate students, who had the longest contribution periods among juniors (with 1.72 years of commit activity compared to 0.98 years for postdocs and 4 months for undergraduates). Similar to senior contributors, junior contributors are significantly involved in creating new features, improving existing capabilities, and fixing bugs.

%\textbf{Graduate students stay longer than postdocs and undergraduate students.} We observed that grad students say, on average, 1.72 years actively contributing to scientific software projects. In comparison, postdocs and undergrads stay, on average, .98 years and 4 months, respectively. Staff members (e.g., professors) are the ones that stay the longer (4.06 years, on average).

%\textbf{Seniors are more active in terms of commits and file creation.} Senior researchers are responsible for a plurality of commits (with an average commit share of 73.21\%), although juniors also contribute substantially (we found one project that has nearly 100\% commits made by junior members). Regarding file creation, \textsf{Senior} contributors tend to be responsible for creating files and, by extension, determining the overall structure of the software project. The ownership of these files are mostly related to those who created the file (i.e., changes are performed in the a ``known'' place).

\item \textbf{An open-source model facilitates external contributions, but the results are mixed.} On one hand, an open-source model makes it easier to attract thirdparty contributors to help grow and maintain the software. However, the software is also made for and by members of a relatively niche and intensely preoccupied community. In the majority of projects we studied, thirdparty contributors tended to be domain expert users who were only active for one day. We also note, however, that these same contributors are more likely to offer defect-correcting commits, which is highly valuable. 

%\textbf{Forward engineering is the most common kind of contribution.}  Half of commits made by postdocs and staff are forward engineering related (43\%, when considering graduate). Collectively (taking sums across all projects), graduate students produce the most forward engineering commits after staff (data?). This suggests that juniors and seniors alike are all responsible for adding new features and capabilities.
%Moreover, 31\% of commits made by thirdparty contributors are corrective (i.e., xxx), which is 2-3$\times$ more frequent than the other roles.

\end{itemize}

\section{Background}\label{sec:bg}

Scientific software projects are very complicated undertakings that have limited budgets, sometimes a lack of software development expertise, and the inherent complexity of the domain~\cite{carver2016software}.

\MyPara{Senior researchers and staff.} As is common in the sciences, a typical software project coalesces around a principal investigator (PI) and one or more co-investigators (Co-Is) who have secured the resources needed for development (e.g. time, money). The reasons for developing software are varied, but include the use-value of the software as a vehicle for research, academic credit, and (in the case of commercial software) revenue~\cite{howison2013incentives}. However, it is well-known that the scientist-as-software-developer rarely has the time to maintain the code that they write~\cite{howison2011scientific}. Time and energy must be divided between writing papers and grant proposals, reviewing manuscripts, mentoring, and conducting experiments \textit{et cetera}. Hannay et al. found while 84\% of interviewees considered developing scientific software important for their own research, the average scientist spent just 30\% of their work time on development activities~\cite{hannay2009scientists}.

%It is well-known that the scientist-as-software-developer rarely has the time to maintain the code that they write~\cite{howison2011scientific}. As noted by Wang et al., scientists ``live and breathe'' their work, with day's labor frequently spilling into night~\cite{wang2012,wang2013exploring}. While scientists enjoy their work, numerous surveys have noted dissatisfaction with excessive workloads~\cite{jacobs2004overworked,mamiseishvili2011examining,fox2011work}. Unfortunately, time and energy must be divided between writing papers and grant proposals, reviewing manuscripts, mentoring, and conducting experiments \textit{et cetera}. Hannay et al. found while 84\% of interviewees considered developing scientific software important for their own research, the average scientist spent just 30\% of their work time on development activities~\cite{hannay2009scientists}. %we should maybe talk about how SE (not scientific developers) also only spend a small fraction of their time on new development (and typically that is only feature implementation) and instead find themselves in many meetings, fixing bugs, etc.

%\textbf{TODO}: Research software engineers (RSEs).

%Hannay et al. 2009 found that researchers typically acquire software development skills (e.g. design, construction) through self-study or through their peers rather than by formal training\cite{hannay2009scientists}.

\MyPara{Junior researchers.} In order to meet the labor needs for development, established researchers often rely upon student and post-graduate labor; juniors are seen as young, full of ideas, and (most importantly) inexpensive personnel~\cite{stephan2012economics}. According to Heroux 2017, senior members provide a stable presence, determining the scientific questions and the trajectory for the software; they are familiar with the conceptual models and the software design, but they may spend less time writing actual code. Juniors, meanwhile, are transient members with a dual focus on contributing code and producing publications; they undergo a staged process of onboarding, becoming experienced, and departing, and during this time they may make substantial contributions to the software~\cite{heroux2017}.

%In an influential study of collaboration, Hara et al. describe the mutually beneficial relationship between senior and junior scientists~\cite{hara2003emerging}. For students, involvement in faculty research projects is an important part of their scientific training as a rite of passage.  Senior researchers benefit both from work \textit{with} juniors (in an advisory role) as well as work \textit{through} juniors (with juniors acting as a ``bridge'' between seniors). However, the character of this collaboration can vary, ranging from complementary participation (each working on discrete units of a project) to integrative participation (joint work on shared components). The projects in this study are spread across that spectrum, with some placing students in a support role and others relying on student contributions to drive the evolution of the core software.

\MyPara{Third party contributors.} For every developer of a critical scientific software package, countless more depend upon it. However, unlike in conventional software development, there are no clear-cut distinctions between users and developers, and, as Turk 2013 argues, trying to force these terms is ``actively harmful'' to our understanding~\cite{turk2013scaling}. Even when they are not directly responsible for a package, it is not uncommon for scientific end-users to write code of their own, such as ``glue code'' that draws together different software tools into a workflow or custom components that utilize others' software. 

The decision to use another's software creates risks because it is not guaranteed that the software will be supported in the future or kept current with the pace of changes in the field~\cite{howison2015understanding}. Converting users into contributors is perhaps even more difficult. However, as Bangherth and Heister 2013 explains, scientific software libraries require the support of a broader community of users and contributors in order to survive in the long term~\cite{bangerth2013makes}. 

%This is, in fact, a common reason for projects to go open-source; several scientific tool papers have variously described open source development as a means for ``nurturing a community'' and promoting an ``active exchange of scientific ideas between users and developers''~\cite{giannozzi2009quantum,oostenveld2011fieldtrip}.

%FieldTrip paper: open source leads to a ``community with an active exchange of scientific ideas between users and developers'', and increases the chances of the development team getting cited for their work\cite{oostenveld2011fieldtrip}.
%Quantum Espresso paper: open source ``nurturing a community engaged in providing [...] solutions''
%OpenSim paper: Making source code available enables researchers to reproduce results produced by other laboratories and to make improvements and adapt code to meet their needs. 

\section{Methodology}

In this section, we describe our research questions and our approach (Section~\ref{sec:rq}). For the repository mining portion of our work, we outline our data collection methodology (Section~\ref{sec:data}), the corpus we have assembled in order to find the answers to our research questions (Section~\ref{sec:projects}), and how we distinguish contributors' roles (Section~\ref{sec:roles}). For our the survey section of this work, we describe our protocol (Section~\ref{sec:surveymethod}).

\subsection{Research Questions}\label{sec:rq}

In this paper, we characterize the habits of scientific software contributors and contributions. Some of our questions are intended to test common wisdom, while others are aimed to probe deeper into the relationships between project contributors. Our research questions are as follows: 

\begin{enumerate}
    \item [\textbf{RQ1:}] What is the tenure of different contributors by role on a scientific software project?
    \item[\textbf{RQ2:}] What is the breakdown of team contributors by role? 
    \item[\textbf{RQ3:}] How much of the development work is done by different contributors by role?
    \item[\textbf{RQ4:}] What kinds of software maintenance and evolution activities do contributors perform?
    \item[\textbf{RQ5:}] How do scientific software developers perceive their own software development process?
\end{enumerate}

Our first question \textbf{RQ1} is demographic in nature based on aggregated project data, and tests the representativeness of our dataset. \textbf{RQ2} enables us to make inferences about the division of labor based on personnel composition. Next, \textbf{RQ3} digs into the kinds of responsibilities, such as file ownership and test files creation, that different contributors take up. \textbf{RQ4} investigates what kind of maintenance and evolution changes, such as adding new features or fixing bugs, do these contributors contribute to the project. Finally, to provide answers to \textbf{RQ5} we surveyed \responses scientific software developers regarding their own contribution behavior.

\subsection{Data Collection Procedure}\label{sec:data}

\begin{figure}[h]
  \centering
  \includegraphics[scale=0.45]{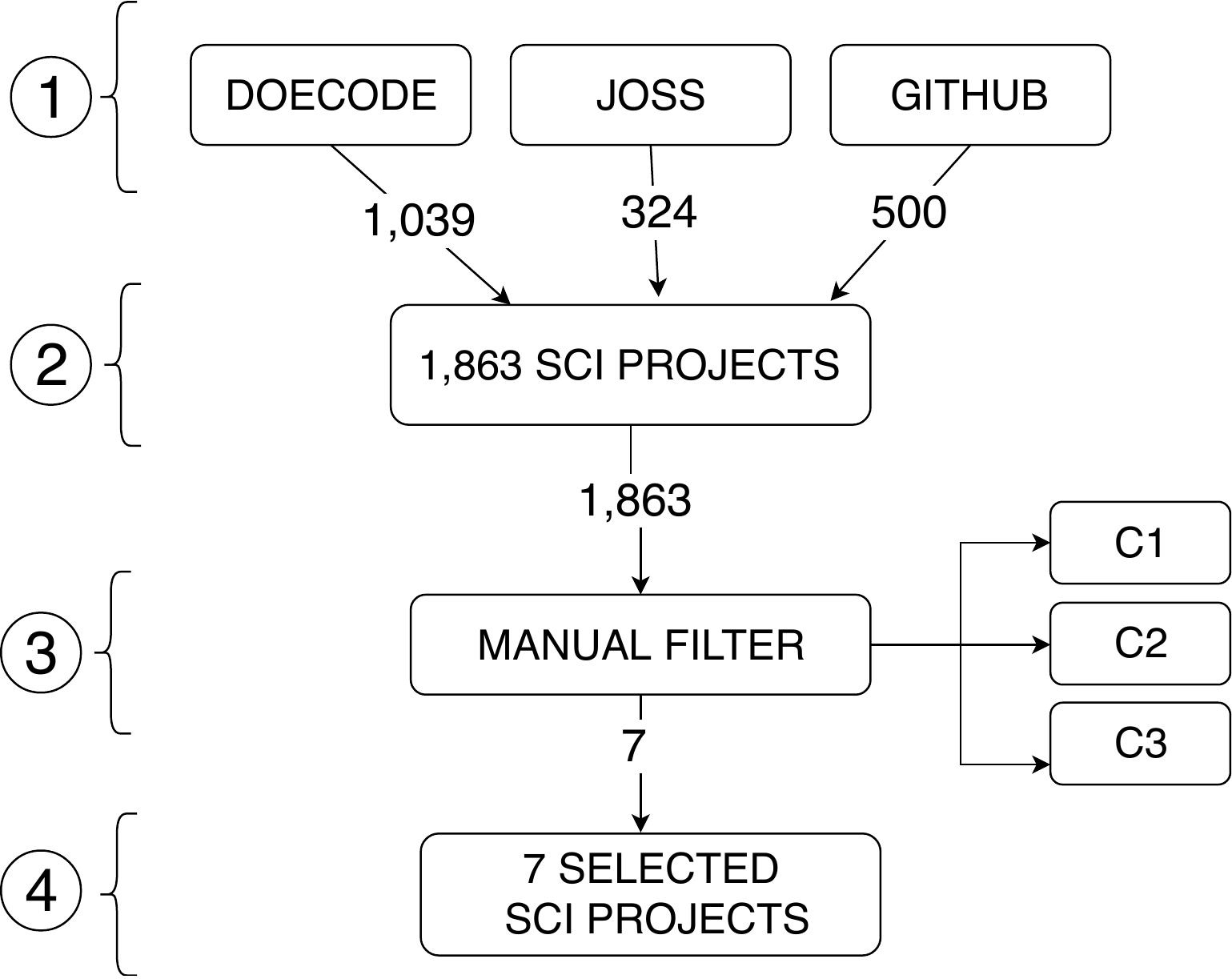}
  \caption{Steps of the data collection procedure.}
  \label{fig:mining-steps}
\end{figure}

Figure~\ref{fig:mining-steps} depicts the steps followed by our data collection procedure.

The first step \circled{1} is aimed to find representative projects. We relied upon three data sources. First, we consulted DOECODE\footnote{https://www.osti.gov/doecode/}, a platform for publicly funded DOE research codes. Next, we searched the Journal of Open Source Software (JOSS)\footnote{https://joss.theoj.org/}, a database of open source research software~\cite{smith2018journal}, which requires all entries be publicly available. Finally, we did searches by topic on GitHub to find repositories with relevant tags (e.g., \texttt{computational-neuroscience}, \texttt{bioinformatics}). This yielded roughly 1,039 repositories from DOECODE, 324 from JOSS, and another 500 from GitHub. These numbers corresponds all projects in these platforms, except for GitHub, in which we stopped searching when we found 500 projects. This resulted in an initial set of 1,863 open source scientific software projects which we chose to take into consideration (step \circled{2}).%\gnote{can we provide a companion website with the projects found, according to each step?}

From this list, we manually analyzed these repositories over several days (step \circled{3}), filtering the results according to the following criteria:

\begin{enumerate}
	\item[C1)] \textbf{Projects should have a contributor list.} The repository must link to a detailed contributor list or research team page that identifies the roles played by different contributors to the project. We use this data to later distinguish contributors' roles (Section~\ref{sec:roles}).
	\item[C2)] \textbf{Projects should be active.} The project must be at least a year old, and the repository must have more than 500 commits. For example, a large number of projects on DOECODE were developed internally and then later released to the public. Thus, the GitHub repository is a shallow copy of the most recent version with no commit history. 
	\item[C3)] \textbf{Projects should be collaborative.} There must be at least three contributors which can be positively identified, and at least one these must be considered a ``junior'' contributor. Many research projects on GitHub are small codes developed by individual researchers in isolation without any significant collaborations with others. Others are collaborative projects between senior staff at different institutions.
\end{enumerate}

After applying these filters, we ended up with a curated list of seven scientific software projects (step \circled{4}).

\subsection{Characterizing the population}\label{sec:pop}

\begin{figure}[ht]
  \includegraphics[width=\linewidth]{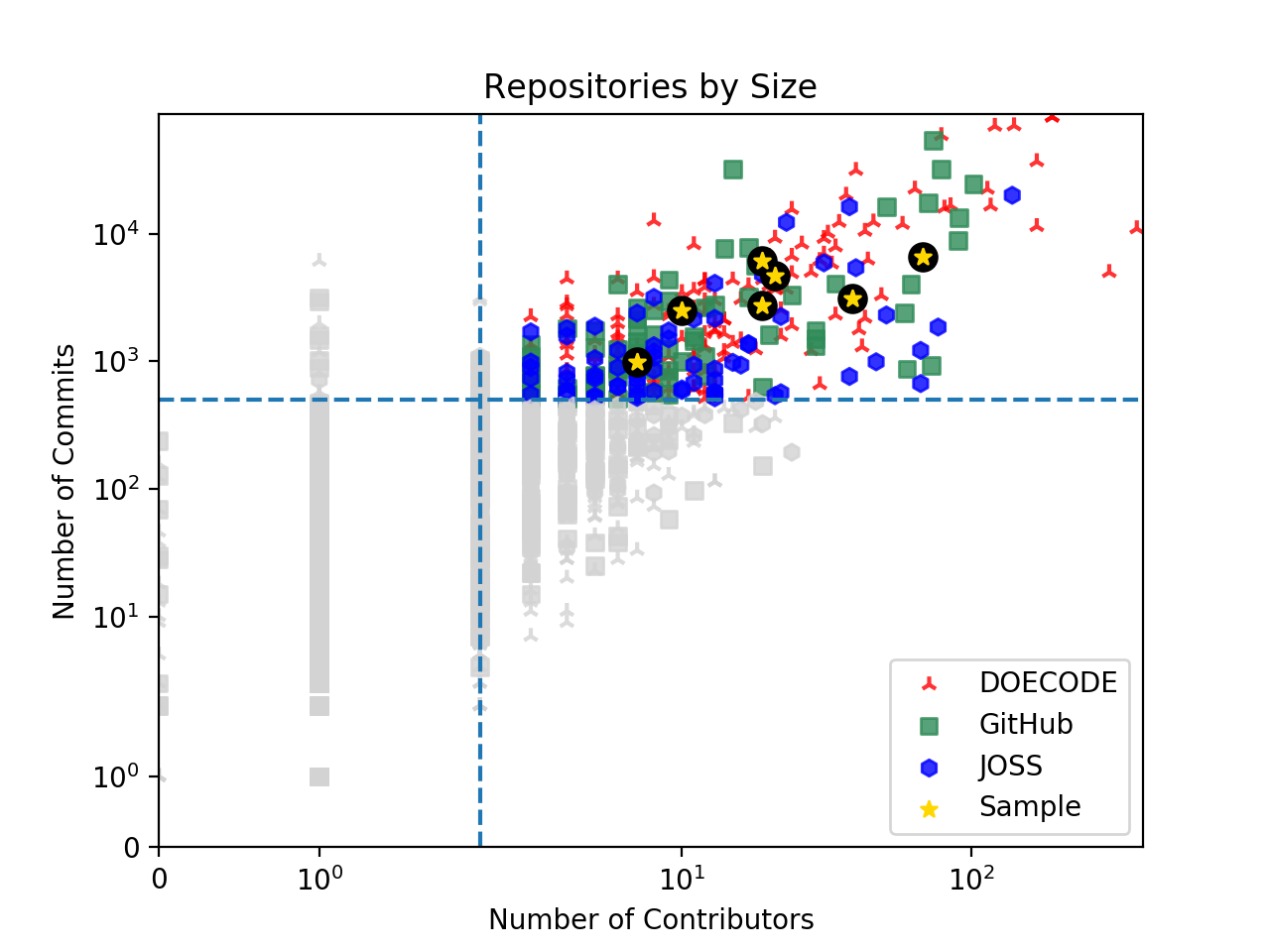}
  \caption{A symmetric log plot of the number of contributors and commits of repositories considered in this work, with those falling beneath the thresholds colored in gray. After these first filtering steps, 359 repositories remain (19\% of the original total), comprising the work of around 10000 contributors.}
  \label{fig:bigpop}
\end{figure}

We believe that the 1,863 projects in our population of repositories we is a representative sample of scientific software projects that can be found in the wild. However, many of these projects are unlikely to provide useful information for our purposes, such as short-lived or single-user research codes, snapshots of codes released for publications, untouched clones of decades-old legacy projects, or mirrors of private repositories lacking history information. As shown on Figure~\ref{fig:bigpop}, filtering for the number of commits and contributors eliminates roughly 8 out 10 of the repositories; the remainder are most likely to be active, collaborative, and (most importantly) to have a rich history on GitHub that we can pull apart. The median project within this group projects has 12 contributors and 1770 commits spread out over 2.93 years.

We can examine a sample of 5000 of the contributors to these repositories, using their number of commits and the number of days spanned by those commits as a proxy for involvement. 29\% of these contributors make only one commit, and 40\% are active for no more than a day. Among contributors more active than these, the average individual has an active tenure of 1.37 years, and during this time they make 116 commits (6\% of the commit activity of the median project). If we zoom in on any one of these contributors, we can observe their activities, but what interests us most is how their identity relates to those activities. This is a more challenging problem to solve, and analysis is much less scalable. Instead, in this work we take a deep dive into a handful of projects whose members we can identify, as a case study into the composition of these teams.

%a variety of different activities, such adding features, fixing bugs, and updating documentation. 

%However, our aim is to pair individuals to 

%However, while this gives us the ``what'', for our work we are also interested in the ``who'' and the ``why'', which, in some respects, are more challenging questions.

\subsection{Studied projects}\label{sec:projects}

Our data collection process yielded a total of seven projects which met our criteria. The descriptions of these projects can be found on Table~\ref{tbl:projectlist}.

\begin{table*}[ht]
\centering
\caption{List of studied projects. Age is present in years. kLOC is calculated using the \texttt{cloc} %\footnote{https://github.com/AlDanial/cloc}
utility, encompassing blank, comments, and code lines. PL means Programming Language.}
\label{tbl:projectlist}
\begin{tabular}{lcrrrclp{4cm}}

Project/GitHub & \begin{tabular}[c]{@{}c@{}}Contributors\\Identified (\%)\end{tabular} & kLOC & Commits & Stars & PL & Age & Description \\ \hline

Chaste (Chaste/Chaste)~\cite{mirams2013chaste} & 97\%  & 371,4k & 4,7k & 22 & C++ & 8 & Tissue and cell level electrophysiology, discrete tissue modeling, and soft tissue modeling \\

Khmer (dib-lab/khmer)~\cite{crusoe2015khmer} & 90\%  & 145,1k & 6.6k & 528  & Python & 7 & Nucleotide k-mer counting, filtering, and graph traversal \\

PyGBe (barbagroup/pygbe)~\cite{cooper2014biomolecular} & 100\% & 12,4k & 0.9k & 28 & Python & 6 & Biomolecular electrostatics and nanoparticle plasmonics \\

LBANN (LLNL/lbann)~\cite{van2015lbann} & 99\%  & 66,8k  & 3,5k & 40 & C++ & 4 & Artificial neural network toolkit \\

Hail (hail-is/hail) & 98\% & 72,9k & 3,1k & 357 & Scala & 2 & Genomic analysis \\

Genn (genn-team/genn)~\cite{yavuz2016genn} & 96\% & 37,4k & 1,8k & 77 & C++ & 6 & Neuron and synapse modeling \\

openMOC (mit-crpg/openMOC) & 90\% & 21,6k & 2,6k & 50 & C++ & 4 & Nuclear reactor physics
\end{tabular}
\end{table*}

Taken together, we argue that our sample of scientific software projects is relevant due to their diversity: 1) they are written in up to four different programming languages (although mainly written in C++, Python, and Scala), with an average of 110.13 kLOC; 2) they span very different domains; 3) they have an average of 5.2 years of historical records; and 4) they are written by scientists that do not necessarily have a computer science or (in particular) a software engineering background. When looking at the number of stars, one might argue that our selected scientific software projects have low popularity. However, as a recent work reported, the median number of stars of R packages published on GitHub is 2~\cite{Pinto:2018:SANER}.

%Khmer describes itself in a paper as an ``open source project focused on sustainable software development'' where they  ``encourage contributions of any kind''. 
%Chaste aims to be open source software because that is a ``pre-requisite for rapid progress, reliability and reproducibility''.
%PyGBe is open source for the sake of reproducibility.

\subsection{Establishing identities of contributors}\label{sec:roles}

Our step was to establish the identities of contributors. We followed the strategy used by Sheltzer and Smith~\cite{sheltzer2014elite} and first scraped data from laboratory websites and project documentation. This was followed up by searching departmental directories and performing web searches in order to disambiguate contributors where necessary. 

The final step in this process is to code each individual in our dataset. This amounts to reviewing the assembled information about each individual and assigning a label to them. For the purposes of this study, we sorted subjects into the following categories: \textsf{undergrad} (i.e., undergraduates students), \textsf{gradstudent} (i.e., master's and doctoral students), \textsf{postdoc} (i.e., postdoctoral researchers), \textsf{staff} (i.e., investigators and support staff), \textsf{thirdparty} (i.e., external collaborators), and \textsf{unknown} (i.e., which the identity could not be established). In cases where we had to rely on incomplete GitHub profile data (e.g. unlisted thirdparty contributors), we attempted to extract names from handles (e.g. \texttt{johnsmith79} $\rightarrow$ \texttt{John Smith}) and cross-referenced those names with web searches for similarly named researchers in the relevant field; where we could not be reasonably convinced that the identities matched, the contributor was left as \textsf{unknown}. To ease understanding, we further group these contributors as \textsf{juniors} (i.e., \textsf{gradstudent}, \textsf{undergrad}, and \textsf{postdoc}), \textsf{seniors} (i.e., \textsf{staff}), and \textsf{thirdparty}.

\subsection{Complementary Survey of Developers}\label{sec:surveymethod}

The majority of the findings of this work come from a quantitative analysis of repositories. To triangulate our findings and better understand the perceptions of scientific software developers, we additionally performed a complementary qualitative analysis of scientific software development teams. We sought to capture, in their own words, (1) who contributes to their projects, (2) how they prepare for contributors to join or leave, and (3) what roles different people play in the development of the software. 

To do this, we designed an online survey. For each participant, we presented three multiple choice questions on their background and experience in developing scientific software, followed by five open-ended questions addressing their team composition and their division of labor. Participation in the survey was voluntary and responses were anonymous. For the open-ended questions, we coded the answers and organized them into categories following the guidelines on open coding procedures~\cite{berg2004methods} (cf. \cite{wood2018detecting}).

To identify the target population, we reached out to development teams whose projects had been accepted to the Journal of Open Source Software (JOSS); we used the JOSS Github repository, which is used to track submissions, to collect information on points-of-contact. From the accepted submissions to JOSS, we identified 273 scientific software developers that either owned or made the majority of contributions to a GitHub project, and recruited them by email. 17 emails were not sent due to mailing errors, and 11 emails were returned due to out of office automatic replies.  Over the period of two weeks, we received \responses answers (a response rate of approximately 30\%).

\section{Analysis}\label{sec:analysis}

%Complementary vs. integrative collaboration.

In this section we provide answers to each research question.

\subsection*{RQ1: What is the tenure of different contributors by role on a project?}

For \textbf{RQ1} we want to know what the expected contribution period is for different contributors based on their project role. While students may spend years with a research team and staff for decades, only a limited portion of that time will be spent on software development activities. Knowing how much labor is available to a team helps staff to understand issues related to task allocation or job rotation. To gather this information, we apply the Kaplan-Meier procedure~\cite{kaplan1958nonparametric} to perform a survival analysis of participants from all projects organized by role. 

In the medical domain, survival analysis measures the fraction of patients who remain alive for a certain amount of time after treatment. In our work, survival refers to how long it takes for a contributor to become inactive. For the purposes of this analysis, we consider a contributor inactive if they have not made a commit within the last 180 days, following the example of Lin et al.~\cite{lin2017developer}; this is more strict than is done in other works (cf.~\cite{nagappan2013diversity}, one commit per year counts as active). We do this because contributors may start or cease their commits in the middle of their tenure (e.g., a junior pivoting towards finishing a thesis). After tuning with different knobs, we found that six months was a reasonable limit. Moreover, recent work has also experimented with different thresholds (e.g., 30, 90, 180 days), and results suggest the same trends over the experiments~\cite{lin2017developer}. We used the \texttt{gitstats} utility\footnote{\url{https://GitHub.com/hoxu/gitstats}} to collect information on the length of each subject's participation in their respective projects.

\begin{figure}[h]
  \includegraphics[width=\linewidth]{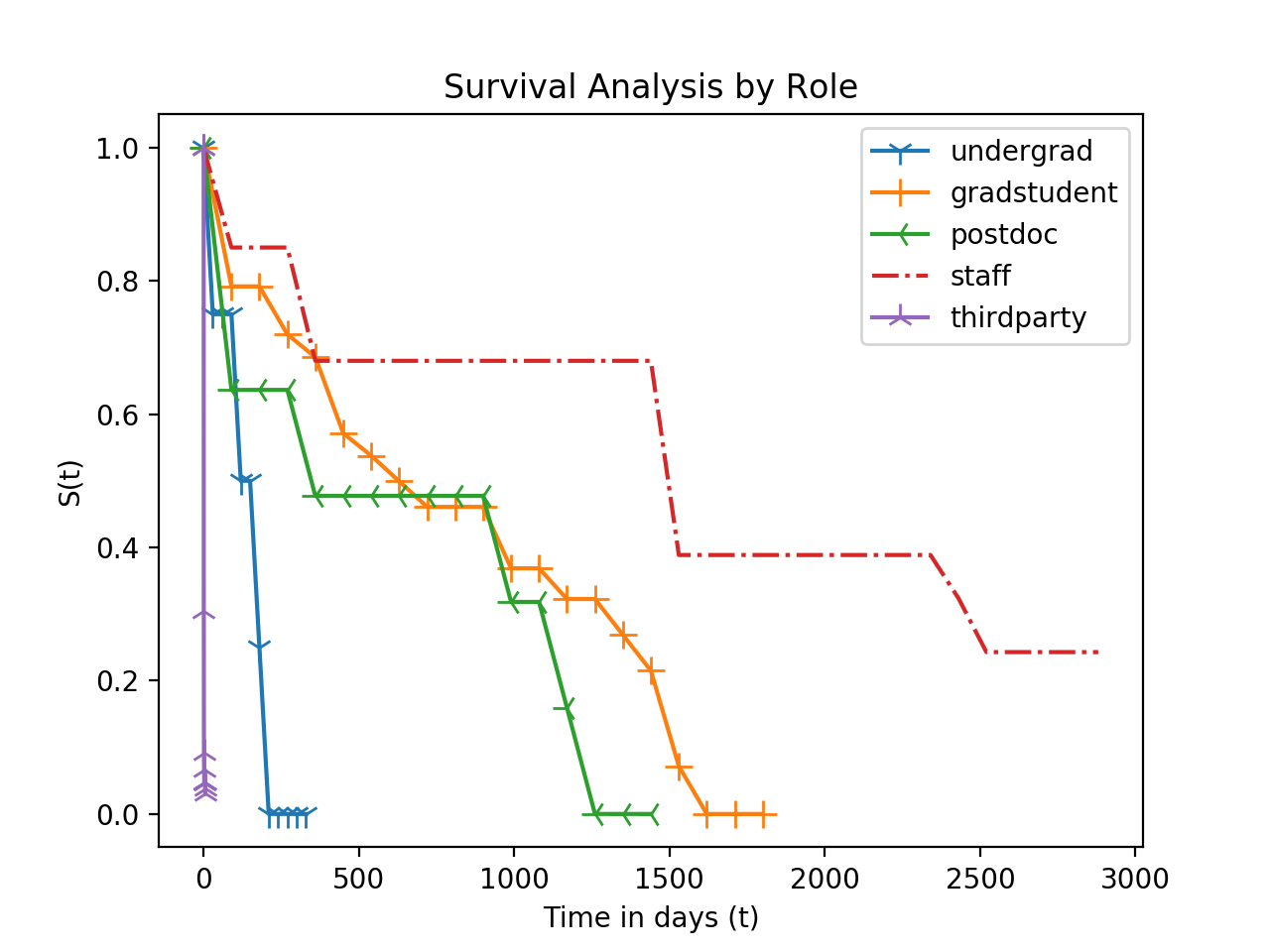}
  \caption{A Kaplan-Meier survival plot of contributors to projects in our dataset, grouped according to role. $S(t)$ indicates the number of individuals in the population who are still actively contributing $t$ days after starting.}
  \label{fig:survival}
\end{figure}

%gradstudent: 630 days
%postdoc: 360 days
%staff: 1485 days
%undergrad: 120 days
%thirdparty: 1 day.
The results of the analysis can be seen on Figure~\ref{fig:survival}. This figure shows a series of declining horizontal steps which approaches the true survival function for that population. The x-axis represents the survival duration, while and the y-axis indicates the probability
that a contributor can survive (i.e., keep actively contributing to the project). The median survival time per group is 4.06 years for \textsf{staff}, 1.72 years for \textsf{gradstudents}, 0.98 years for \textsf{postdocs}, 4 months for \textsf{undergrads}, and just a day for \textsf{thirdparty} contributors. 

%\gnote{can we break down this per project? or even better, can we have one figure per project?}

A careful, contextualized reading of this data proves informative. With regards to \textsf{juniors}, aside from \textsf{staff}, we see that \textsf{gradstudents} are the most likely to stay around. That being said, a typical PhD student (as almost all of the graduate students in our dataset are) takes 5 years to graduate, and this means that the median \textsf{gradstudent} only spends 34.52\% of their tenure contributing to a project. Meanwhile, \textsf{postdocs} spend even less time than \textsf{gradstudents} as contributors, even though the typical term limit of \textsf{postdocs} today is also 5 years~\cite{ferguson2014national}; that being said, \textsf{postdocs} are brought on-board with prior knowledge that they can immediately apply to the software project. Finally, \textsf{undergraduate} students in our dataset only spend one semester out of their 4 year education participating in developing scientific software. Taken all together, the median \textsf{junior} in our projects spends only 24.84\% of their time in their position doing software development work.

Meanwhile, \textsf{seniors} provide the most stable presence, with a median survival time of 4 years. We note that this is affected by right-censoring because the average age of our projects is 5 years. However, our evidence suggests that \textsf{senior} members who do contribute code may not do so indefinitely. Once the software reaches a point of maturity, they may hand off the work to \textsf{juniors}. In other cases, they may leave virtually all the work to \textsf{juniors}. Lastly, we found that in the projects we studied, \textsf{thirdparty} contributors tend to remain at the periphery and do not engage with a project for any significant length of time; as we observed, \textsf{thirdparty} contributors stay, on average, 1 day.

\subsection*{RQ2: What is the breakdown of team contributors by role?}

For \textbf{RQ2}, we are interested in the breakdown of contributors to each project by role. How many people contribute to projects overall? How can we characterize them? A key distinction that we are making is that this is not the same thing as the number of contributors listed as team members on the webpage of the project or research group. For example, a graduate student may be a user of the software but not a contributor, and in the case of projects like Chaste, we can have multiple staff members responsible for design work and guidance of students but who have no commits to their name.

%There are two ways that we can do this, which we can phrase as subquestions: 

%\begin{itemize}
%	\item \textbf{RQ2.1}: How many people (grouped by role) contribute to projects overall?
%    \item \textbf{RQ2.2}: How many different people (again, grouped by %role) contribute to the project concurrently?
%\end{itemize}

\begin{figure}[h]
  \centering
  \includegraphics[scale=0.37]{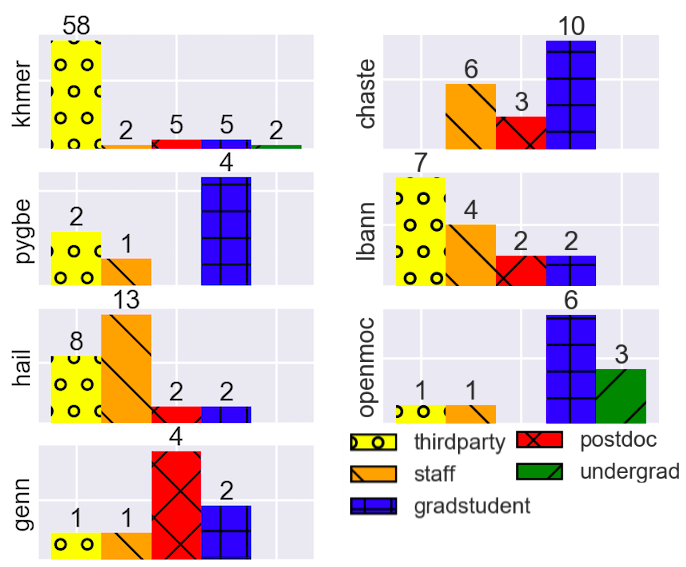}
  \caption{For \textbf{RQ2}, for each project we present a bar chart with totals of identified contributors sorted by role.}
\label{fig:population}
\end{figure}

We showcase our results on Figure~\ref{fig:population}. Taken together, \textsc{junior} members make up the majority of team contributors (average 69\%; median 80\%), with the remainder being \textsc{seniors} (average 31\%; median 20\%). Meanwhile, in all but one of the projects we studied, we were able to identify \textsf{thirdparty} contributors (average 20\%; median 27\%). Additionally, as a rule, both team members and \textsf{thirdparty} contributors that we identified have a background relevant to the domain of the project, something which we learned by analyzing biographical information used to classify contributors by role.

%\MyBox{\textbf{RQ2}: \textsf{Juniors} make up the majority of team members on most studied projects, followed by \textsf{seniors}. In all but one project, we identify at least one \textsf{thirdparty} contributor. As a rule, identified contributors had a background relevant to the domain of the software project.}

%\begin{figure}
%\begin{subfigure}[b]{0.45\textwidth}
%   \includegraphics[width=1\linewidth]%{figs/contributionPeriods.png}
%   \label{fig:Ng1} 
%\end{subfigure}

%\begin{subfigure}[b]{0.45\textwidth}
%   \includegraphics[width=1\linewidth]%{figs/contributionPeriods.png}
%   \label{fig:Ng2}
%\end{subfigure}
%\caption{Above, a sample of overlapping contribution periods for different team members on the Khmer project. Below, a PLACEHOLDER.}
%\end{figure}

\subsection*{RQ3: How much of the development work is done by different contributors by role?}

%For \textbf{RQ3}, we want to characterize the amount of development work that is done by these different actors. %This question is somewhat vague because there are many different kinds of work that go into a project and different ways in which those activities can be measured. Instead, 
%We can refine our question into more concrete subquestions:

%\begin{itemize}
%	\item \textbf{RQ3.1}: Who makes the most commits?
%    \item \textbf{RQ3.2}: Who creates the most files?
%    \item \textbf{RQ3.3}: Who creates the most test files?
%    \item \textbf{RQ3.4}: What is the interplay between who edits a code file and who creates the code file?
%\end{itemize}

%\begin{figure*}
%  \includegraphics[width=\textwidth,height=4cm]{figs/barcharts.png}
%  \caption{The results gathered in service of \textbf{RQ3}. Each bar represents a kind of contributors, varying from \textsf{juniors}, \textsf{seniors}, to \textsf{thirdparty}. The first line answers \textbf{RQ3.1}, while the second answers \textbf{RQ3.2}, and the third answers \textbf{RQ3.3.}}
%  \label{fig:contribuions}
%\end{figure*}

%\merge{Figure~\ref{fig:contribuions} shows the overall results for this section.} For \textbf{RQ3.1}, the number and frequency of commits are well-worn metrics for engagement and investment in a project~\cite{nagappan2013diversity,hattori2008nature}. 
For \textbf{RQ3}, we want to characterize the amount of development work that is done by these different actors.  To begin, we consider the share of commits produced by different contributors, the number and frequency of commits being well-worn metrics for engagement and investment in a project~\cite{nagappan2013diversity,hattori2008nature}. 
Taking averages across all projects that we studied, half or more of commits are made by \textsf{senior} members (average 50.76\%; median 63.29\%). However, the majority of the other half are commits by \textsf{junior} contributors (average 42.9\%; median 36.7\%). Moreover, for 4 out of 7 projects (Khmer, Chaste, Hail, and openMOC), \textsf{senior} researchers are responsible for a plurality of commits (with an average commit share of 77.12\%; median 73.34\%); the opposite is true for Genn, LBANN, and PyGBe (average share 23.92\%; median 22.58\%), with LBANN having a more even split between \textsf{junior} and \textsf{senior} members.  We also note that, for the projects we have studied, \textsf{thirdparty} contributors tend to play a very minor role in this regard; only LBANN has a notable share of commits attributable to \textsf{thirdparty} users (17.7\% versus an average of 3.59\% and median of 0.06\%). Overall, in all projects but one (Hail), \textsf{junior} contributors produce a significant share of commits, meaning that even when \textsf{seniors} do most of the heavy lifting, \textsf{juniors} play an essential role in the realization of the project.

%One reason that might explain this behavior is due to the unique arrangement provided by large research institutions such as national labs.

However, while both \textsf{junior} and \textsf{senior} alike generate significant amounts of commit activity, this is not to say that the \textit{scope} of their activities is comparable. To better understand this, we consider interactions with and ownership of files. For the purposes of this work, we record a user as interacting with a file each time they make a commit that touches that file. By that measure, the typical \textsf{junior} interacts with a much smaller percentage of files compared to a \textsf{senior} (average 5.85\% vs. 20.35\%; median 0.65\% vs. 11.51\%). This is to say that a distinguishing characteristic of \textsf{junior} developers in our corpus is that they often have a narrow focus on a particular subset of a project. Meanwhile, the same is especially true for \textsf{thirdparty} contributors who interact with an even smaller percentage of files (average 1.64\%; median 0.66\%).

Likewise, we can also consider file creation. Earlier work by Poncin et al.~\cite{poncin2011process} addresses file creation in their operationalization of ``core'' developers, as frequent creation and modification of files indicates that a user is helping to drive the vision or direction of the software. Related to this, in a recent study of large-scale open source projects, Lin et. al~\cite{lin2017developer} found users who created files tended to be longer-term contributors than those who modified files. In 5 out of 7 of the projects we studied (Khmer, Chaste, Hail, openMOC, and Genn), \textsf{senior} team members created the majority of files (average of 69.44\%); LBANN is almost evenly split by this measure, and PyGBe, as a student-driven project, has only a quarter of its files originating from \textsf{senior} members. 

\subsection*{RQ4: What kinds of maintenance and evolution activities do contributors perform?}

What value do different kinds of contributors add to a project? For example, once a \textsf{gradstudent} exits a project, in what ways did they influence the evolution of the software during their tenure? We can find some evidence for this through project pages and documentation when teams provide an itemized list of accomplishments of different contributors (as is the case for several of the projects in our study); it is typical to see \textsf{juniors} receiving credit for implementing novel features pertinent to their research, \textsf{seniors} for building out infrastructure and performing maintenance, and \textsf{thirdparty} contributors for providing support or helping improve the codebase. However, relatively few projects provide this kind of fine-grained information, and we would also like to be able to interrogate those claims in an empirical way.

To answer our question, we present two views of the development activities that elucidate the kinds of work that different contributors produce. The first is an analysis of commit messages based on the approach of Hattori and Lanza~\cite{hattori2008nature}, and the second is an analysis of file paths involved in commits based on the work of Vasilescu et al.~\cite{vasilescu2014variation}. In both cases, we are interested in categorizing commit activity according to their purpose or intent. 

In the framework set out by Hattori and Lanza~\cite{hattori2008nature}, commits are divided into four major categories of activity: 

\begin{enumerate}
\item \textbf{Forward engineering (Fwd)}, for instance, adding new features;
\item \textbf{Reengineering (Reng)}, for instance, refactoring activities;
\item \textbf{Corrective (Corr)}, for instance, fixing bugs;
\item \textbf{Management (Mgmt)}, for instance, updating documentation.
\end{enumerate}

In order to automatically classify commits into these categories, the authors compare the content of commit messages against predefined word banks for each commit type based on the earliest match found. For instance, consider the following commit message: ``\emph{This commit adds integrators supporting the combined, staggered, and pseudotransient forward sensitivity analysis methods where the sensitivity equations are solved alongside the forward equations}.''\footnote{\url{https://GitHub.com/trilinos/Trilinos/commit/e8e6d67}} Unpacking the semantics of this commit message requires extensive domain knowledge and that is difficult to automate. However, the word ``add'' is a match in the word bank for forward engineering; it is reasonable to assume (in this case) that the commit is adding a new feature to the software.

%\gnote{can we add the complete list of word bank?}

This approach is limited in that it only considers the lexical content of messages, and it also fails to handle situations where a commit may belong to more than one category, but it remains useful as a diagnostic tool. To test the validity of this classification scheme against our corpus, we chose to manually classify a representative random sample of commits drawn from across all projects. Assuming that all projects have statistically similar commits (in the sense that the distribution of commits by type are roughly the same), a sample of 378 commits might reflect the overall population of roughly 23,000 commits with a confidence level of $95\%$ with an interval of $\pm5\%$. In order to arrive at the ground truth, our manual classification considers not only commit messages but also the artifacts (such as source code, documentation, and test data) that were modified and the context in which that occurred (such as preceding commits and related files). 

\begin{table}[h]
\centering
\caption{Confusion matrix for validation of Hattori-Lanza~\cite{hattori2008nature} classification scheme. Unknown (Unk) commits are those which the algorithm failed to classify.}
\label{tbl:htconfusion}
\begin{tabular}{llllll}
\multicolumn{1}{l|}{\diagbox{Automated}{Manual}} & Fwd & Reng & Corr & Mgmt                    &  \\ \cline{1-5}

\multicolumn{1}{l|}{Fwd}                             & \cellcolor{gray!25} 53  & 11   & 1    & \multicolumn{1}{l|}{14} &  \\
\multicolumn{1}{l|}{Reng}                            & 3   & \cellcolor{gray!25} 60   & 0    & \multicolumn{1}{l|}{13} &  \\
\multicolumn{1}{l|}{Corr}                            & 1   & 4    & \cellcolor{gray!25} 60   & \multicolumn{1}{l|}{5}  &  \\
\multicolumn{1}{l|}{Mgmt}                            & 7   & 14   & 11   & \multicolumn{1}{l|}{\cellcolor{gray!25} 16} &  \\ \cline{2-5}
Unk                                                  & 9   & 36   & 5    & 55                      & 
\end{tabular}
\end{table}

Table~\ref{tbl:htconfusion} shows the confusion matrix between the automated approach and the manual analysis. First, we note that because a commit may fail to match against the word bank, it is possible for this approach to fail to find a label. This happened for 27\% of commits in our sample. We identified three causes for this: (1) manual classification relied on words that were outside of the word bank (e.g., ``\textit{vectorized} book-keeping kernels''), (2) commit messages were automatically generated and vague as to their purpose (e.g., merging an arbitrary pull request), and (3) messages could be highly ambiguous (e.g., ``complete breakdown of intuition'' or ``it is all becoming clear''). However, we consider this to be a more gentle form of failure than attempting to shoehorn an unintelligible commit into an arbitrary category.

For commits which were classified automatically, the manual and automatic approaches agreed 69\% of the time; much of the error was concentrated on management commits, only a third of which were correctly labeled. If we limit our consideration to just forward engineering, reengineering, and corrective commits, then the automatic approach agrees 89\% of the time, with some minor confusion between forward and reengineering activities. As such, we limit our consideration to just those three. 

The second approach we use is derived from that of Vasilescu et al.~\cite{vasilescu2014variation}, which categorizes commit activity by examining the filepaths involved in changes; file extensions (e.g. \texttt{.cpp} versus \texttt{.csv}) and hints in file paths (e.g. \texttt{/src/} versus \texttt{/test/} can clue us in to the purpose of a file and, by extension, the kind of labor that an individual provides a project through their interaction with those files. The classification algorithm itself is analogous to what was previously described: filepaths are matched against a bank of regular expressions that map to different categories of files. Unlike with the Hattori-Lanza scheme, these results are much less ambiguous because it is reasonable to assume that file extensions indicate actual file types. For this work, we made several addenda to the regexes used in the original paper in order to cover additional programming languages (e.g., Pascal and Ada), data storage types commonly used in scientific computing (e.g., HDF5 and FASTA), as well as a handful of previously unaddressed build and configuration artifacts (e.g., Dockerfiles and Gradle build files). On Table~\ref{tbl:commitshares} and Table~\ref{tbl:fileactivities} we provide results for our two analyses as an aggregate of all contributors in the projects we studied as a way of approximating a ``typical'' project. The former shows what percentage of commits made by an average individual are categorized as forward engineering, reengineering, and corrective activities; the latter asks what percentage of individuals have made at least one commit that interacts with a file of a given type. 

As a group, \textsf{senior} contributors have the highest average share of forward engineering commits (33.37\%), which is to say that commits made by seniors are more likely to include novel development work, such as adding or extending software capabilities. \textsf{Senior} developers also play a key role in realizing the supporting infrastructure of their projects, with a majority having interactions with \texttt{build}, \texttt{devdoc}, and \texttt{metadata} files. Likewise, a plurality of \textsf{seniors} interacts with \texttt{data/database} files (such as data for validation tests and parameters for research models) and \texttt{test} files.%, both of which have special significance in this domain. 

%\MyBox{\textbf{RQ4}: \textsf{Senior} contributors frequently engage in forward engineering activities, and are most likely to interact with the build system, documentation, testing, and data files.}

Next, the activities of \textsf{junior} contributors resemble that of \textsf{senior} contributors in many key respects. Like \textsf{seniors}, \textsf{juniors} universally interact with \texttt{code} files, and a comparable share of their commits go towards forward engineering (26.77\% for \textsf{juniors} vs. 33.37\% for \textsf{seniors}), reengineering (22.11\% vs. 18.90\%) , and corrective  (15.93\% vs. 14.91\%) activities. Also like \textsf{seniors}, the majority of \textsf{juniors} interact with \texttt{build}, \texttt{devdocs}, and \texttt{test} files (though in smaller measures compared to \textsf{seniors}), and this was true in general for all projects we studied. This reinforces our earlier observations suggesting that their work is neither subordinate nor peripheral compared to the work of \textsf{seniors}, but is instead vitally important to the enterprise.

%\MyBox{\textbf{RQ4}: Behaviorally, \textsf{juniors} largely resemble \textsf{seniors} in creating new features, improving existing code, and fixing defects. The majority of \textsf{juniors} make commits that touch \texttt{build}, \texttt{devdocs}, and \texttt{test} files, which indicate an interest in maintaining and improving the codebase.}

Finally, we consider \textsf{thirdparty} contributors who, as we determined earlier, are relatively minor players who as a rule only sporadically contribute to projects. \texttt{Code} and \texttt{devdocs} aside, they scarcely interact with any kind of file. One point that stands out, however, is that these contributors are more likely to make corrective activities (with an average commit share of 35.06\%). This is to say that while they make very few commits, the commits they do make are more likely to be bug fixes; that suggests that \textsf{thirdparty} contributors are most likely to be users of the software who have the domain knowledge and development expertise needed to correct such bugs or ``scratch their own itch''. In essence, \textsf{thirdparty} contributor behavior is similar to the kind of work produced by peripheral developers, which are typically involved in bug fixes, and they have irregular or short-term involvement in a project~\cite{Crowston:2006:HICSS,Pinto:SANER:2016}.

%\MyBox{\textbf{RQ4}: \textsf{thirdparty} contributors primarily interact with \texttt{code} artifacts with a strong interest in fixing and improving the code that they rely upon. In that scientific software must be trustworthy to be useful, this makes them very valuable.}

\begin{table}[]
\centering
\caption{The relative share of automatically classified commits of an average junior, senior, or thirdparty contributor. Management commits are omitted.}
\label{tbl:commitshares}
\begin{tabular}{llll}
                                   & Fwd Share & Reng Share & Corr Share \\ \cline{2-4} 
\multicolumn{1}{l|}{Juniors}       & 26.76\%                   & 21.11\%             & 15.93\%          \\
\multicolumn{1}{l|}{Seniors}       & 33.37\%                   & 16.07\%             & 14.91\%          \\
\multicolumn{1}{l|}{thirdparty} & 19.98\%                   & 18.89\%             & 39.45\% 
\end{tabular}
\vspace{-0.2cm}
\end{table}

%\begin{figure}
%\begin{subfigure}[b]{0.45\textwidth}
%   \includegraphics[width=1\linewidth,scale=0.35]{figs/forward.png}
%   %\label{fig:Ng1} 
%\end{subfigure}
%\begin{subfigure}[b]{0.45\textwidth}
%   \includegraphics[width=1\linewidth,scale=0.35]%{figs/reengineering.png}
   %\label{fig:Ng2}
%\end{subfigure}
%\begin{subfigure}[b]{0.45\textwidth}
%   \includegraphics[width=1\linewidth,scale=0.35]{figs/corrective.png}
%   %\label{fig:Ng2}
%\end{subfigure}
%\label{fig:commitshares}
%\caption{The relative share of automatically classified commits of an average junior, senior, or thirdparty contributor. Senior contributors are more likely to create forward engineering commits, and thirdparty contributors are more likely to create corrective commits.}
%\end{figure}

%\cite{vasilescu2014variation} does not do multiple classification either.

\begin{table}[b!]
\vspace{-0.2cm}
\centering
\caption{The percentage of contributors in each category who have at least one commit that interacts with a given file type. N/A indicates that no matches were found for regexes in any projects studied for a given category}
\label{tbl:fileactivities}
\begin{tabular}{llll}
                                      & Juniors & Seniors & thirdparty \\ \cline{2-4} 
\multicolumn{1}{l|}{Documentation}    & 19.1\%  & 26.0\%  & 0\%           \\
\multicolumn{1}{l|}{Images}           & 14.5\%  & 22.2\%  & 2.7\%         \\
\multicolumn{1}{l|}{Localization}     & 2\%     & 0\%     & 0\%           \\
\multicolumn{1}{l|}{UI}               & N/A     & N/A     & N/A           \\
\multicolumn{1}{l|}{Media}            & 27.7\%  & 33.3\%  & 2.7\%         \\
\multicolumn{1}{l|}{Code}             & 100\%   & 100\%   & 70.3\%        \\
\multicolumn{1}{l|}{Project Metadata} & 36.2\%  & 51.85\% & 8.1\%         \\
\multicolumn{1}{l|}{Configuration}    & 34.0\%  & 33.3\%  & 5.4\%         \\
\multicolumn{1}{l|}{Build}            & 63.8\%  & 77.8\%  & 29.7\%        \\
\multicolumn{1}{l|}{Devdocs}          & 63.8\%  & 88.8\%  & 66.7\%        \\
\multicolumn{1}{l|}{Data/Databases}   & 36.2\%  & 63.0\%  & 18.9\%        \\
\multicolumn{1}{l|}{Test}             & 74.5\%  & 85.2\%  & 27.0\%        \\
\multicolumn{1}{l|}{Libraries}        & N/A     & N/A     & N/A          
\end{tabular}
\end{table}

\subsection*{RQ5: How do scientific software developers perceive their own software development process?}

For our final research question, we consider how scientific software developers view their own projects based on our survey data; this provides us with points of comparison with our quantitative findings. Among our survey respondents, those we identified as being most responsible for their respective projects, 36\% of them were postdocs, 30\% were non-academic professionals, another 30\% were students (undergraduate or graduate), and 14\% were professors. The majority of them (62\%) work for an university or college, 11\% work for the government, 8.5\% work in industry, and the other 18.5\% play other roles. Regarding their highest academic degree, 76.4\% had already received or were working towards their doctorate degree (18\% have a master degree, and only 4\% have bachelor degree). They worked in a variety of fields, including computer science, fine art, chemistry, political science, urban planning, and neuroscience. 

The majority of projects in our survey (61\%) were developed by a team of people, though it is worth nothing that a significant number of projects were the work of individuals (39\%). On average, these teams had 3.6 contributors (3rd Quartile: 4.2, max: 15); this roughly aligns with the number of active contributors in a given year in the repositories which we mined (average: 3.8, 3rd quartile: 5, max: 11). 

When we asked (Q5) what and how do they contribute, we observed that 50 respondents reported software development-oriented activities such as fixing bugs, developing scripts to support research, improving documentation, and adding tests. Strongly tied with software development activities, 18 respondents reported to contribute to non-software activities, such as paper writing, grant writing, running experiments, etc. Along these lines, three respondents perceive their contributing role as ``\emph{Conducting research that feeds back into the project}''. Regarding how do scientific software developers get trained to do their jobs (Q6), 70\% of the respondents were self-taught, although some of them received mentorship from senior contributors (e.g., ``Shadowing a more senior developer for a week or two''), while others benefited from online training programs (e.g., ``The Molecular Sciences Software Institute (molssi.org) training programs.''), took advantage of their own documentation (e.g., ``We make sure that the docs are self-contained to ease onboarding for remote teams''), or even the pull-request process (e.g., ``By first contributing some pull requests and getting code reviews''). Only four respondents were trained through their academic degree.

%The scientific software projects in our survey were mostly composed by a single person (40\% of them develop software alone); on average, teams have 3.6 contributors (3rd Quartile: 4.2, max: 15).

When considering the responsibilities they need to take to prepare for the departure of a team member (Q7), 20 respondents mentioned the importance to keep the documentation updated (e.g., ``We simply try to ensure that all developments are adequately documented at the time, to help the understanding of future developers''). Interestingly, 8 respondents mentioned that this never happened, which is partially because they are working on a small or solo team (e.g., ``No one has departed yet (or joined...)''). Other respondents mentioned the need to train other, to push code, or to add tests.

Of those projects run by teams by teams of two or more people, 35 out of 41 specifically called attention to the role played by \textsf{juniors} in developing their software. 20 of these described them as being responsible for developing specific, non-core features of the software; projects that followed this pattern offered up explanations such as ``[juniors] --- by necessity -- start with smaller peripheral bugs and features. Core development requires a lot of experience and knowledge.'' Another twelve projects, however, cast \textsf{juniors} as being developers of core infrastructure, typically for the reasoning that \textsf{senior} members ``cannot afford to put much time into development''.

Meanwhile, 24 of the 41 team projects emphasize the role of \textsf{seniors} in development. 8 of these said seniors developed the core of the software and 4 the periphery. Those that did so often emphasized the need for experience in development, insofar as ``the more education a team member has (software development life-cycle, good coding practices, etc.), the better they are at seeing `big picture' development tasks [...] these people often lead development''. However, in contrast to this, ten respondents characterized \textsf{seniors} as being visionaries first and developers second. In this view, the role of \textsf{seniors} is to ``coordinate activities'', ``drive the direction of the project'', ``guide the conceptual development'', and to provide the ``theoretical details''.

Lastly, only 9 out of the \responses projects gave recognition to \textsf{thirdparty} contributors. Among these projects, the typical view was that while \textsf{thirdparties} ``contribute seldomly'', they were also a common sources of bug fixes, a finding echoed in our findings from \textbf{RQ4}. Likewise, these contributors were also responsible for ``[submitting] small patches to make [a] tool better meet their own niche use cases''.

%in this sense, while \textsf{thirdparty} contributors have only a very ephemeral presence in our data, they are likely

\section{Discussion}\label{sec:discussion}

\begin{table}[]
\caption{A summary with descriptions of typical contributors, based on the findings in this work.}
\begin{tabular}{ll}
Contributor & Findings \\ \hline
\multicolumn{1}{l|}{Seniors} &  
\begin{minipage} [t] {0.35\textwidth} 
\begin{itemize}
    \item Are often active on a project for four or more years.
    \item Make the majority of contributions to a typical project (average 50\%). They create the most files and touch the most code.
    \item Are most often responsible for forward engineering activities, development of the core of the software, and infrastructure tasks.
    \item Provide guidance and visionary leadership to junior contributors, especially when they do not have the time or resources to work on the code themselves.
\end{itemize}
\end{minipage} \smallskip
\\ \hline
\multicolumn{1}{l|}{Juniors} &  
\begin{minipage} [t] {0.35\textwidth}
\begin{itemize}
    \item Are active for no more than 2 years. Roughly 25\%-35\% of their time is spent on software development.
    \item Make up the majority of team members.
    \item Perform many of the same development activities as seniors, and, collectively, generate 42\% of commit activity on average. 
    \item Are most likely to develop peripheral, non-core features of the software. 
\end{itemize} \smallskip
\end{minipage}

\\ \hline
\multicolumn{1}{l|}{Third Parties} & 
\begin{minipage} [t] {0.35\textwidth}
\begin{itemize}
    \item Are active for only one day.
    \item Have a background in the domain of the project and an interest in using the software.
    \item Make only a handful of commits. These commits are most likely to be bug fixes. 
\end{itemize}
\end{minipage} \smallskip
\end{tabular}
\label{tbl:summary}
\vspace{-0.3cm}
\end{table}

We have summarized the major findings in Table~\ref{tbl:summary}, and now consider the potential implications of our work.  

\textbf{Training.} As a group, \textsf{juniors} have long been the subject of science public policy literature. Novice researchers are ``canaries in the mine'' for the health of the scientific enterprise, as it is during this period that they are meant to learn the values and skills needed to participate fully as scientists~\cite{louis2007becoming}. However, while software development is an increasingly important skill, the amount of direct experience they acquire may be limited by competing demands in their academic careers (see \textbf{RQ1}). This brings into focus a number of different topics regarding software sustainability (e.g., the importance of good practices such as code reviews) as well as educational policies (e.g. the need for more formal software development training in STEM curricula).

\textbf{Building Communities.} Much has been written about the value of openness in science and the need for community support of scientific software. As we noted in section ~\ref{sec:pop}, 40\% of contributors stay on for only a day. Our analysis suggests that many of these may be \textsf{thirdparty} users who have a formal background in a domain relevant to the project; even though these users only be involved for a short time, they can still make valuable contributions such as fixing bugs. However, despite the widespread presence of short-term committers in the population and \textsf{thirdparty} contributors being present in all but one of the projects in our sample, only 12\% of respondents in our companion survey mentioned these contributors. Based on this, we believe that better community policies could help attract these contributors, such as providing guides for new contributors and explicitly giving credit to these users.

\textbf{Supporting Sustainability.} Scientific software projects are known for being long-lived and under constant pressure to keep pace with scientific advances. It is common for \textsf{senior} project members to provide visionary leadership to guide that process, but how this translated to software construction was unclear. Our findings place \textsf{seniors} in a primary role as core developers who are most likely responsible for forward engineering and infrastructure tasks. Our findings suggest directions for future research, such as how research priorities generate software development tasks and when and how those tasks are delegated. A more complete understanding of this phenomenon would help software engineers develop better tools and techniques to support that effort.

\section{Threats to Validity}

%We must recognize that there are multiple caveats with our approach to find project members.
First, not all members of a project show up as contributors to the repository; for example, senior staff may offer guidance and support while leaving the actual implementation work to others. Second, people who stop contributing code may still be part the project. This is frequently the case for graduate student contributors, who may refocus on completing coursework or a thesis towards the end of their tenure. Third, team websites are not always up-to-date, and not all contributors are given explicit recognition for their work; in one case, an undergraduate student was uncredited on a project site, but a subsequent web search uncovered a university press release detailing a research grant that was awarded for them to do specific work on the project in question. Finally, not all contributors are project members; as with most open-source software projects, open-source scientific software attracts third-party contributors, typically senior researchers who benefit from using the software.

%this isn't open coding...maybe we need to introduce the term coding in the paper.  for example.. In this paper, we use the open coding methodology.  Open coding is the practice of labeling items based on.....It is commonly used in sociolgy, anthropology, etc.  
Coding individuals using our approach means addressing several potential ambiguities. First, on a few occasions, a subject may have belonged to different categories at different times (e.g., a staff member starting off as a postdoc). When this occurred, we labeled them according to the role in which they made the majority of their commits. Second, for large research institutions (e.g., national labs), a software project may receive contributions from people nominally part of the same organization, but unaffiliated with the research team; we resolved this by treating those contributors as \textsf{thirdparty} contributors.  

The commit analysis performed to answer \textbf{RQ4} was an automated process which, when applied to a large-scale number of commits, can silently yield false-positives (i.e., commits that were unable to be categorized), since commit messages might lack semantical sense~\cite{Maalej:2010:MSR} or are even empty~\cite{Dyer:2013:ICSE} (i.e., zero words). 
To mitigate such bias, we conducted a manual analysis over a representative sample of 378 commits (confidence level of $95\%$ with an interval of $\pm5\%$). We observed a low number of uncategorized commits. Although uncategorized commits still exist, we believe this approach is the fairest way to categorize the commit intention because, since we are dealing with scientific software projects, the domain of our studied projects is highly specific and complex. Therefore, any other attempt to categorize commits using our own domain experience would introduce even more bias.

Lastly, one could argue that this study does not provide a novel contribution, e.g., ``obviously graduate students are largely responsible for adding new features''. However, such common-sense assumptions are often not backed up by empirical evidence. This paper piles such evidence and, more importantly, quantifies the phenomenon; even though some perceptions are confirmed (e.g., ``\textsf{gradstudents} stay longer than \textsf{postdocs} and \textsf{undergraduates}''), other are uncovered (e.g., ``\textsf{thirdparty} members survive only one day on average'').

\section{Related Work}

%In this section we describe the studies overlapping with the scope of our work.

\noindent
\textbf{Studies on scientific software development.} %Several recent studies have considered the social aspects of collaboration and software engineering practice in the context of scientific software development. 
Turk~\cite{turk2013scaling} presents techniques for encouraging community engagement with scientific software in the astrophysics community, arguing that attracting thirdparty contributors requires intentional actions designed to encourage their participation. Likewise, Bangherth and Heister~\cite{bangerth2013makes} outlines the practices of successful open-source scientific software libraries, which includes a discussion of the value proposition behind open-sourcing software primarily written by juniors. Sletholt et al.~\cite{sletholt2011literature} performs a case study of agile development practices among scientific software projects. Howison and Hersleb~\cite{howison2011scientific}, a qualitative study of the incentives for creating and maintaining scientific software, identifies the general breakdown of contributors to projects that they study; however, both the focus and methods used in that work differ from ours, as they are not concerned with the specifics of how different people contribute to the software development. Finally, Storer~\cite{storer2017bridging} provides an excellent survey of the state of software engineering practice within the scientific community. Our work shares the same motivations as others, but, to our knowledge, is the first to tackle this subject from a software repository mining perspective.

\noindent
\textbf{Studies on roles of contributors in open source projects.} % There is a recent flow of work aimed at characterizing the different roles played by contributors in open source projects. 
The study of core developers, i.e., developers that play an essential role in developing the system architecture and forming the general leadership structure, in open source projects is a fruitful research area~\cite{Fielding:1999:CACM,Licorish:2014:IST,Coelho:2018:CHASE}. Core developers are well-known from being active contributors. A general rule of thumb suggests that contributors with more than 80\% of the overall contributions are considered core developers~\cite{Crowston:2006:HICSS,Mockus:2002:TOSEM}. Indeed, for some projects, this number is even higher. Recent work indicates that several well-known, active open source projects rely on 1--2 core developers to drive most of their maintenance and evolution tasks~\cite{Avelino:2016:ICPC}. On the periphery side, research indicates that peripheral developers accounts for more than 90\% of the contributors of a project~\cite{Crowston:2006:HICSS}. %One particular kind of peripheral developer is the so called casual contributors (also called as drive-by commits or one-timers), which are developers that contribute just once, and then leave the project. 
Moreover, several authors have acknowledged the existence and the growth of casual contributors (i.e., developers that contribute just once)~\cite{Pham:2013:ICSE,Pinto:SANER:2016,Lee:2017:ICSE}. Here we enriched the understanding of core and peripheral developers.
We also introduced the notion of third-party contributors, which share some of the behaviors commonly observed in peripheral developers (e.g., they have a short term relationship with the project, and most of their contributions are intended to fix bugs).

%We observed that staff members are in charge of driving the project, they also ... \gnote{do staff members made more than 80\% of the commits?} 

\section{Conclusion}

Scientific software projects are critical to the advancement of the scientific enterprise, and software engineering research can directly help those efforts through tailored tools, techniques, and practices. However, there has historically been a lack of hard data on who contributes to scientific software and how they behave. In this work, we mined logs from seven non-trivial open source scientific software projects in order to provide answers to these questions. Among our findings, we found that while \textsf{senior} researchers perform the lion's share of the work in many projects, \textsf{junior} researchers are often on the frontlines driving the software development. We also considered the habits of \textsf{thirdparty} contributors who, while often operating at the periphery of projects, have a valuable role to play in fixing and improving code.

%As we outlined in our Section~\ref{sec:discussion}, our analysis provides a foundation for future research activities. 
For future work, we plan to conduct ethnographic studies of scientific software projects to better understand topics such as feature creation and bug fixing. We also plan to study how scientific software projects compare with conventional projects.

\vspace{0.2cm}
\noindent
\emph{Acknowledgments.} We thank the reviewers, the \responses respondents, and PROPESP/UFPA and CNPq (\#406308/2016-0).

%We also plan to investigate abandoned open-source scientific software projects.%, since they could derive some insights for development by comparing successful/failure projects.

%Different than, say, regular software projects, the development of scientific software projects is mainly driven by academics, which might be involved for a short term (e.g., students working in a project as part of their requirements) or long term (e.g., professors that use the software to explore new research paths). However, it so far unclear how does these contributors behave, in term of activity (how long do they stay?), responsibility (who drive the project?), or workability (what do they do?).

%In this paper we mined logs from seven non-trivial open source scientific software projects in order to provide answer these questions. Among the findings, we observed that juniors (e.g., graduate students) do xxx. On the other hand, seniors (e.g., professors) do yyy. 

% This paper motivates us to interview more teams. We were very restricted in what kinds of projects we could target because we needed to have all the information on staffing readily available.

%\subsection{Future Work}

%More work is needed to understand the needs and aims of third-party contributors (they do not have an affiliation with the project, although they want to contribute, nevertheless). We also plan to conduct interviews with juniors and seniors members in order to cross-validate the findings of our study.

\bibliographystyle{IEEEtran}
\bibliography{references}

\section{Appendix}
\begin{table}[H]
\caption{Addenda to File Extensions used in Vasilescu et al.\cite{vasilescu2014variation}}
\begin{tabular}{lp{6cm}}
Category & Addenda \\ \hline
\multicolumn{1}{l|}{Doc} &
\begin{minipage} [t] {0.85\linewidth}
\begin{verbatim}
".*\.md"
\end{verbatim}
\end{minipage}
\\ \hline
\multicolumn{1}{l|}{Code} &
\begin{minipage} [t] {0.80\linewidth}
\begin{verbatim}
".*\.pas((\.swp)?)(~?)",
".*\.pxd((\.swp)?)(~?)",
".*\.ads((\.swp)?)(~?)",
".*\.adb((\.swp)?)(~?)",
".*\.bin"
\end{verbatim}
\end{minipage}
\\ \hline
\multicolumn{1}{l|}{Devdoc} &
\begin{minipage} [t] {0.80\linewidth}
\begin{verbatim}
".*\.pdf",".*citation.*",
".*license.*",".*doxyfile.*",
".*\.wiki",".*\.tex",
".*\.bib",".*\.dox",".*authors"
\end{verbatim}
\end{minipage}
\\ \hline
\multicolumn{1}{l|}{Db} &
\begin{minipage} [t] {0.80\linewidth}
\begin{verbatim}
".*\.csv",".*\.xml",".*\.fa",
".*\.xlsx",".*\.zip",".*\.h5",
".*\.bz2",".*\.tar(\.gz)?",
".*\.fq(\.gz)?",".*\.pts",
".*\.pdb",".*\.pqr",
".*\.vert",".*\.node",
".*\.edge",".*\.param(eters)?",
".*\.phi0",".*\.prototext(\.bve)?",
".*\.pkl",".*\.pbs"
\end{verbatim}
\end{minipage}
\\ \hline
\multicolumn{1}{l|}{Build} &
\begin{minipage} [t] {0.80\linewidth}
\begin{verbatim}
".*\.build",".*dockerfile",
".*\.gradle"
\end{verbatim}
\end{minipage}
\\ \hline
\multicolumn{1}{l|}{Config} &
\begin{minipage} [t] {0.80\linewidth}
\begin{verbatim}
".*\.vcxproj((\.filters)?)(~?)",
".*\.qpg",".*\.dsp",".*\.epf"
\end{verbatim}
\end{minipage}
\\ \hline
\multicolumn{1}{l|}{Img} &
\begin{minipage} [t] {0.80\linewidth}
\begin{verbatim}
".*\.graffle"
\end{verbatim}
\end{minipage}
\end{tabular}
\end{table}

\subsection{Questions Used in the Online Survey}
\newcommand{\surveyquestion}[1]{
%\begin{tabular}
%\begin{minipage}[t] {0.85\linewidth}
{
\begin{minipage}{0.85\linewidth}
#1
\end{minipage}\smallskip \\
}
%\end{minipage}
%\end{tabular}
}

\surveyquestion{
1) Which of the following best describes you (select all that apply)?
\begin{itemize}
    \item Student (e.g. undergraduate or graduate student)
    \item Postdoc
    \item Professional
    \item Professor
\end{itemize}
}
\surveyquestion{
2) If you are currently employed, which of the following best describes your current employer?
\begin{itemize}
    \item Government
    \item University or college
    \item Business or industry
    \item Non-profit organization
    \item Other (please specify)
\end{itemize}
}
\surveyquestion{
3) Please list the highest academic degree you have received or that you are currently working toward.
\begin{itemize}
    \item High school degree or equivalent
    \item Associate's Degree
    \item Bachelor's
    \item Master's
    \item Doctorate
\end{itemize}
}
\surveyquestion{
4) What is the subject of this degree?
}
\surveyquestion{
5) How many people are on your team?
}
\surveyquestion{
6) What and how do they contribute (e.g., adding new features to the core of the project, conducting research that feeds back into the project, writing tests, fixing bugs, et cetera.)?
}
\surveyquestion{
7) How does a new team member get trained to do their job (e.g. they are self-taught)?
}
\surveyquestion{
8) What (if anything) do you do to prepare for the departure of a team member (e.g. we ask them to document the undocumented parts of their code)?
}
\surveyquestion{
9) What kinds of roles and responsibilities do different people play in the development of your software? In particular, consider the following groups: junior researchers (undergraduates, graduate students, and postdocs), senior researchers (seasoned staff), and third party contributors.
}
\surveyquestion{
10) Are there differences in what different people provide your project? If so, please explain what (e.g. juniors are responsible for small issues while seniors drive the architecture and solve the main problems)?
}

\end{document}